# A Weekly Pattern from Hourly Estimates of the Super-Kamiokande-I Neutrino Flux 1996-2001

Lasse E. Bergman
SIS, University of California, San Francisco
San Francisco, California 94143-0404

September 19, 2005

## Abstract

A search for a neutrino flux difference between weekdays and weekend days, for the average week of the Super-Kamiokande-I (SK-I) Experiment, was undertaken using the 5-day period version of the SK-I data taken from May $31^{st}$, 1996 to July $15^{th}$, 2001. A significant ($p \ll 0.001$) difference was found and the most obvious neutrino flux change from weekdays to weekend days can be summarized as follows: "Some neutrinos took the weekend off – especially on Saturday".

## I. INTRODUCTION

The Super-Kamiokande Collaboration [1,2,3] has placed neutrino data sets in the public domain [4]. The author used one of those data sets in order to estimate the hourly Neutrino Flux for its average week. The Collaboration has not found any seasonal or short-time fluctuations in the SK-I neutrino flux for these data sets (see below).

The purpose of this paper was to search for even shorter short-time fluctuations, i.e. for a weekly pattern showing a neutrino flux difference between weekdays and weekend days, as suggested by the results in [5]. And as in [5], the particular data set used in this paper was "collected at SK from May 31st, 1996 to July 15th, 2001, yielding a total detector live time of 1,496 days. This data taking period is known as SK-I", yielding some 15 events per day i.e. approximately 22,400 neutrino events for 1996-2001. The data set is the one arranged as 5-day periods, i.e. "neutrino data, acquired over 1,871 elapsed days from the beginning of data-taking, … divided into roughly … [5-day] long samples as listed in Table … " ( TABLE II in [4] ).

## II.  NO PERIODIC VARIATIONS REPORTED BY THE COLLABORATION

Rather than exploring daily changes, the Collaboration searched for seasonal and short-time fluctuations but did not find any significant periodicity in the SK-I neutrino flux [4], thus ruling out "semiannual (seasonal) variations of the observed solar neutrino flux because of the changing magnetic field caused by the 7.25 degree inclination of solar axis with respect to the ecliptic plane" and any "short-time variation … due to the 27-day rotation of the Sun".  So, with the exception of the Homestake experiment [7], "Kamiokande and other experiments have not provided any evidence for a time variation of the neutrino flux outside of statistical fluctuations  [6] "  [4].

## III.  HOURLY ESTIMATES

By necessity, the neutrino flux numbers presented in this paper are estimates, and these estimates leave something to be desired, for the following reasons.

First, hourly comparisons based on raw data would require access to un-binned experimental data, e.g. the date and time for each individual neutrino event.  However, there are good reasons why no event-by-event summary of the SK-I data is yet publicly available.  As the author was graciously informed, the rationale for not releasing un-binned SK-I data publicly might roughly be expressed as follows: "At the event level, interpretation of systematic errors, calibrations, and background subtraction, gets quite complicated, such that for someone not close to the gory details of the detector and reconstruction software, doing a proper job gets very difficult.   This seems to be a common problem in high energy physics, unlike astronomy where raw data is routinely made public." [8].  Consequently, the 358 binned neutrino flux values in Table II in [4] were published with statistical uncertainties (for example 2.74 +0.63 -0.53 $10^6$ cm$^{-2}$ s$^{-1}$, for the first 5-day period in Table II in [4] ).  Fortunately, the random effects of this statistical variation upon the published flux values tend to cancel out each other when some 260 (Table I, below) of those 358 values are included in calculating the numerator of each weighted neutrino flux mean.

Second, "There are on and off periods of data-taking in the [5-day period] and thus the timing of each [5-day period] is calculated as the mean of the start and end times and corrected by SK livetime. Hence the mean time is not necessarily an exact division of the time interval for each [5-day period]." [4]   In this paper, as in [5], the unknown (for the author) off periods of data-taking within the 5-day periods by necessity had to be ignored.   This ignorance had implications for interpreting the results in Figure 1, below.

## IV.  RESULTS

The results in this paper are visible in Figure 1 (next page), where each of the 168 "hourly estimates" from Table I (below) has been plotted against its corresponding Day ( MON, TUE, WED, THU, FRI, SAT, or SUN ) and Hour ( "0:00-0:59", … , or "23:00-23:59" ).  Figure 1 suggests that neutrino flux changes occurred from weekdays to weekend days.  The most obvious change can be summarized as follows: "Some neutrinos took the weekend off – especially on Saturday".  Figure 1 also displays local maxima at most (5 out of 7) midnights (cf. [5] ).

The 120 (5*24=120) weighted weekday means and the 48 (2*24=48) weighted weekend day means in Figure 1 and in Table I may also be viewed as two samples consisting of 120 and 48 plain values, respectively.  Statistically, one can calculate the likelihood that these two samples were randomly drawn from identical populations, e.g. the likelihood (the p-value) that the difference between the two sample means was caused by chance alone.  A (two-sided) two-sample t-test [9] for the difference between the two population means $\mu_{WeekdayHr}$ and $\mu_{WeekendHr}$ produced a significant
( $p = 7.32 * 10^{-16} \ll 0.001$ ) result, giving a strong reason to reject the null hypothesis
$H_o: \mu_{WeekdayHr} = \mu_{WeekendHr}$ .

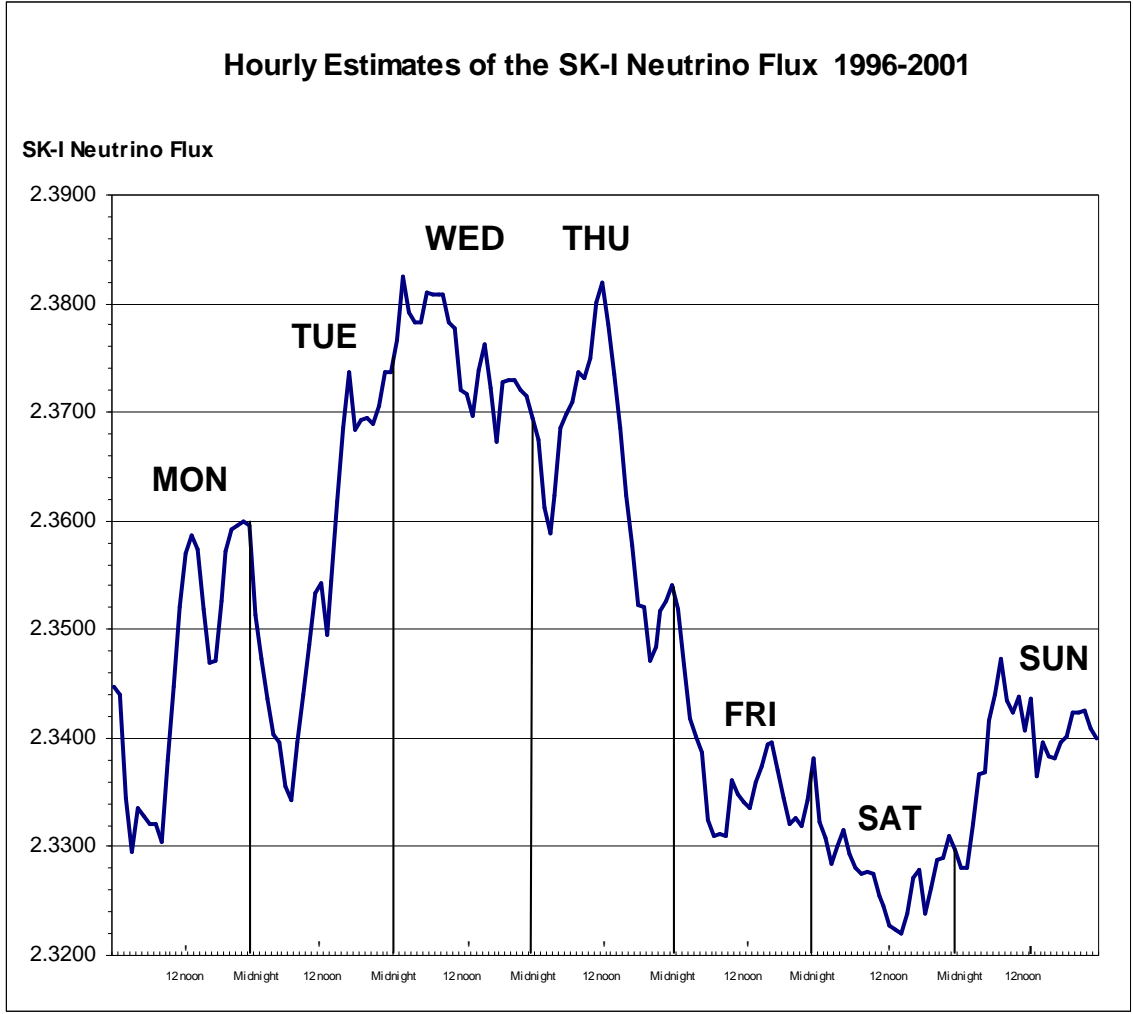

FIGURE 1: Hourly Estimates of the SK-I Neutrino Flux for the Average Week of 1996-2001. The horizontal axis is the day of the average week, and within each day are its 24 hours 0:00-0:59, 1:00-1:59, … , 23:00-23:59. The vertical axis is the weighted hourly estimate of the SK-I neutrino flux, for a particular day and hour. A weight (out of about 260 weights for that particular day and hour) is the minutes (typically 60 minutes) covered by that particular day and hour during a 5-day period in Table II in [4] and the neutrino flux value weighted by that weight is the SK-I neutrino flux for that particular 5-day period in Table II in [4]. The vertical axis has the units of $10^6$ cm$^{-2}$ s$^{-1}$. Please see Table I and the text for more details.

The sample means were $\bar{y}_{WeekdayHr} = 2.3556$ and $\bar{y}_{WeekendHr} = 2.3335$, with the sample standard deviations $s_{WeekdayHr} = 0.016526$ and $s_{WeekendHr} = 0.007282$, and the sample sizes $n_{WeekdayHr} = 120$ and $n_{WeekendHr} = 48$. This significant result suggests that the neutrino flux was different between weekdays and weekend days.

Figure 1 is thus based on the 168 neutrino flux estimates in Table I ( 7 columns * 24 rows = 168 estimates) and each of those estimates is a weighted mean. Each term in the numerator of the weighted mean is a product of a neutrino flux value (the neutrino flux of a particular 5-day period) and a time weight (typically 60 minutes during that 5-day period). In Table I the number within parentheses, after the weighted mean, denotes how many products were used to calculate that particular mean.

For example, the first product (out of 264 products) for the column "FRI" and the row "4:00-4:59" in Table I has a time weight of 29 minutes and a neutrino flux value of 2.74 ( FRIDAY 05/31/1996 4.31 thru 4:59 in 5-day period No. 1 in Table II in [4] where the neutrino flux for this first 5-day period is 2.74 * $10^6$ $cm^{-2}$ $s^{-1}$). To exemplify further, the second product ( out of 264 products) for the column "FRI" and the row "4:00-4:59" in Table I has a time weight of 60 minutes and a neutrino flux value of 2.83 ( FRIDAY 06/07/1996 4:00 thru 4:59 in 5-day period No. 2 in Table II in [4] where the neutrino flux for this second 5-day period is 2.83 * $10^6$ $cm^{-2}$ $s^{-1}$). ). And, the third product ( out of 264 products) has a time weight of 60 minutes and a neutrino flux value of 2.30 ( FRIDAY 06/14/1996 4:00 thru 4:59). And so on. The "hourly estimate" for the column "FRI" and the row "4:00-4:59" in Table I was then calculated as the weighted mean ( 29*2.74 + 60*2.83 + 60*2.30 + … ) / ( 29 + 60 + 60 + … ) = 2.3388 , in units of $10^6$ $cm^{-2}$ $s^{-1}$ .

## V. DISCUSSION

As noted above, the off periods of data-taking within the 5-day periods in Table II in [4] had to be ignored but they have implications for the interpretation of the results in Figure 1. As mentioned, a (two-sided) two-sample t-test for the difference between the two means $\mu_{WeekdayHr}$ and $\mu_{WeekendHr}$ produced a significant ( p = 7.32 * $10^{-16}$ << 0.001 ) result. But such a result would have to be interpreted with caution, since the 120 and 48 values of the two samples respectively were weighted neutrino flux means that were calculated with (typically) 60-minute weights. And some of these 60-minute weights included such off periods of data-taking.

Of course, the calculated difference between the two sample means $\bar{y}_{WeekdayHr}$ and $\bar{y}_{WeekendHr}$ depends on which 5-day periods are included in the calculations. For instance, a split-half strategy to test the consistency for the whole 1996-2001 period produced a (less) significant ( $p = 1.045 * 10^{-3} \sim 0.001$ ) result when only the 179 odd-numbered 5-day periods (periods 1, 3, … , 357) were included, but a more significant ( $p = 4.51 * 10^{-35} \ll 0.001$ ) result when only the 179 even-numbered 5-day periods (periods 2, 4, … , 358) were included. As another consistency test, a more significant ( $p = 4.24 * 10^{-31} \ll 0.001$ ) result followed when 5-day periods for only the year of 1998 (the periods numbered 111-182) were included (cf. [5]).

Being outside the scope of this paper, the 10-day period version of the SK-I data in Table I in [4] was not included in the calculations mentioned above. Such 10-day periods (only 184 10-day periods vs. 358 5-day periods) would not have resulted in a significant difference between weekdays and weekend days, but rather a significant ( $p = 7.72 * 10^{-4} < 0.001$ ) difference between "short week Mon-Thu" days and "long weekend Fri-Sun" days.

For a possible reason why a neutrino flux difference between weekdays and weekend days should be expected, please see [5].

_______________________________________

Table I: Hourly Estimates of the Super-Kamiokande (SK-I) Neutrino Flux for the average week of 1996-2001. Each estimate is a <u>weighted mean</u>. Each term in the numerator of the weighted mean is a product of a <u>neutrino flux value</u> (for a particular 5-day period in Table II in [4]) and a <u>time weight</u> (typically 60 minutes for the particular day and hour in that 5-day period). The number within parentheses, after the weighted mean, denotes how many products were used to calculate that particular mean. The neutrino flux estimates have the units of $10^6$ cm$^{-2}$ s$^{-1}$. Please see Figure 1, and the text for more details.

| 24 HOURS PER DAY | **MON** | | **TUE** | | **WED** | | **THU** | | **FRI** | | **SAT** | | **SUN** | |
|---|---|---|---|---|---|---|---|---|---|---|---|---|---|---|
| | Neutrino Flux Weighted Mean | Number of Weights | Neutrino Flux Weighted Mean | Number of Weights | Neutrino Flux Weighted Mean | Number of Weights | Neutrino Flux Weighted Mean | Number of Weights | Neutrino Flux Weighted Mean | Number of Weights | Neutrino Flux Weighted Mean | Number of Weights | Neutrino Flux Weighted Mean | Number of Weights |
| 0:00- 0:59 | **2.3448** | (266) | **2.3514** | (268) | **2.3767** | (265) | **2.3674** | (264) | **2.3519** | (266) | **2.3323** | (266) | **2.3280** | (263) |
| 1:00- 1:59 | **2.3441** | (269) | **2.3472** | (264) | **2.3825** | (264) | **2.3613** | (263) | **2.3468** | (263) | **2.3307** | (266) | **2.3280** | (263) |
| 2:00- 2:59 | **2.3345** | (266) | **2.3437** | (262) | **2.3791** | (264) | **2.3588** | (265) | **2.3419** | (262) | **2.3284** | (263) | **2.3321** | (267) |
| 3:00- 3:59 | **2.3295** | (266) | **2.3404** | (264) | **2.3782** | (266) | **2.3624** | (267) | **2.3402** | (262) | **2.3301** | (262) | **2.3366** | (264) |
| 4:00- 4:59 | **2.3336** | (265) | **2.3396** | (264) | **2.3783** | (266) | **2.3686** | (264) | **2.3388** | (264) | **2.3315** | (263) | **2.3369** | (264) |
| 5:00- 5:59 | **2.3329** | (265) | **2.3355** | (262) | **2.3811** | (265) | **2.3698** | (264) | **2.3325** | (264) | **2.3293** | (261) | **2.3416** | (267) |
| 6:00- 6:59 | **2.3321** | (263) | **2.3343** | (264) | **2.3808** | (263) | **2.3709** | (264) | **2.3310** | (265) | **2.3280** | (261) | **2.3441** | (265) |
| 7:00- 7:59 | **2.3321** | (263) | **2.3396** | (263) | **2.3809** | (266) | **2.3736** | (264) | **2.3312** | (264) | **2.3276** | (261) | **2.3473** | (265) |
| 8:00- 8:59 | **2.3304** | (265) | **2.3440** | (266) | **2.3808** | (265) | **2.3732** | (265) | **2.3309** | (268) | **2.3277** | (264) | **2.3435** | (266) |
| 9:00- 9:59 | **2.3380** | (264) | **2.3486** | (262) | **2.3782** | (261) | **2.3749** | (265) | **2.3361** | (264) | **2.3276** | (263) | **2.3424** | (264) |
| 10:00-10:59 | **2.3447** | (258) | **2.3534** | (261) | **2.3777** | (262) | **2.3802** | (263) | **2.3349** | (262) | **2.3254** | (262) | **2.3438** | (265) |
| 11:00-11:59 | **2.3521** | (258) | **2.3542** | (262) | **2.3721** | (260) | **2.3820** | (259) | **2.3342** | (260) | **2.3245** | (264) | **2.3407** | (263) |
| 12:00-12:59 | **2.3570** | (258) | **2.3495** | (260) | **2.3717** | (260) | **2.3780** | (262) | **2.3335** | (262) | **2.3228** | (259) | **2.3436** | (266) |
| 13:00-13:59 | **2.3586** | (259) | **2.3544** | (259) | **2.3696** | (259) | **2.3734** | (264) | **2.3360** | (262) | **2.3223** | (261) | **2.3365** | (265) |
| 14:00-14:59 | **2.3575** | (257) | **2.3617** | (262) | **2.3739** | (265) | **2.3685** | (266) | **2.3374** | (265) | **2.3220** | (258) | **2.3397** | (264) |
| 15:00-15:59 | **2.3518** | (260) | **2.3685** | (263) | **2.3763** | (260) | **2.3623** | (263) | **2.3394** | (265) | **2.3239** | (260) | **2.3382** | (263) |
| 16:00-16:59 | **2.3470** | (262) | **2.3736** | (258) | **2.3722** | (262) | **2.3577** | (266) | **2.3397** | (261) | **2.3272** | (259) | **2.3381** | (264) |
| 17:00-17:59 | **2.3471** | (259) | **2.3684** | (260) | **2.3674** | (262) | **2.3522** | (261) | **2.3370** | (262) | **2.3278** | (261) | **2.3397** | (264) |
| 18:00-18:59 | **2.3526** | (260) | **2.3692** | (262) | **2.3727** | (263) | **2.3521** | (264) | **2.3345** | (261) | **2.3238** | (263) | **2.3401** | (265) |
| 19:00-19:59 | **2.3573** | (259) | **2.3695** | (262) | **2.3730** | (262) | **2.3471** | (262) | **2.3320** | (262) | **2.3262** | (261) | **2.3423** | (263) |
| 20:00-20:59 | **2.3592** | (261) | **2.3688** | (263) | **2.3730** | (262) | **2.3483** | (264) | **2.3326** | (264) | **2.3288** | (264) | **2.3424** | (264) |
| 21:00-21:59 | **2.3595** | (262) | **2.3706** | (263) | **2.3720** | (264) | **2.3517** | (264) | **2.3319** | (263) | **2.3290** | (263) | **2.3425** | (264) |
| 22:00-22:59 | **2.3600** | (259) | **2.3737** | (262) | **2.3715** | (262) | **2.3525** | (263) | **2.3344** | (266) | **2.3309** | (266) | **2.3409** | (267) |
| 23:00-23:59 | **2.3597** | (260) | **2.3737** | (262) | **2.3695** | (265) | **2.3541** | (264) | **2.3382** | (265) | **2.3298** | (265) | **2.3400** | (264) |